
\catcode`@=11

\def\singlespace{\normalbaselines}
\def\oneandahalfspace{\baselineskip=1.15\normalbaselineskip plus 1pt
\lineskip=2pt\lineskiplimit=1pt}

\def\np{\vfill\eject}
\def\nl{\hfil\break}

\def\nofirstpagenoten{\nopagenumbers\footline={\ifnum\pageno>1\tenrm
\hss\folio\hss\fi}}
\def\nofirstpagenotwelve{\nopagenumbers\footline={\ifnum\pageno>1\twelverm
\hss\folio\hss\fi}}
\def\leaderfill{\leaders\hbox to 1em{\hss.\hss}\hfill}
\def\ft#1#2{{\textstyle{{#1}\over{#2}}}}
\def\frac#1/#2{\leavevmode\kern.1em
\raise.5ex\hbox{\the\scriptfont0 #1}\kern-.1em/\kern-.15em
\lower.25ex\hbox{\the\scriptfont0 #2}}
\def\sfrac#1/#2{\leavevmode\kern.1em
\raise.5ex\hbox{\the\scriptscriptfont0 #1}\kern-.1em/\kern-.15em
\lower.25ex\hbox{\the\scriptscriptfont0 #2}}

\parindent=20pt
\def\narrow{\advance\leftskip by 40pt \advance\rightskip by 40pt}

\def\AB{\bigskip
        \centerline{\bf ABSTRACT}\medskip\narrow}
\def\nonarrower{\advance\leftskip by -40pt\advance\rightskip by -40pt}
\def\AE{\bigskip\nonarrower}

\def\underscore#1{$\underline{\vphantom{y}\hbox{#1}}$}

\def\boxit#1{\vbox{\hrule\hbox{\vrule\kern3pt
        \vbox{\kern3pt#1\kern3pt}\kern3pt\vrule}\hrule}}

\def\gtorder{\mathrel{\raise.3ex\hbox{$>$}\mkern-14mu
             \lower0.6ex\hbox{$\sim$}}}
\def\ltorder{\mathrel{\raise.3ex\hbox{$<$}|mkern-14mu
             \lower0.6ex\hbox{\sim$}}}
\def\dalemb#1#2{{\vbox{\hrule height .#2pt
        \hbox{\vrule width.#2pt height#1pt \kern#1pt
                \vrule width.#2pt}
        \hrule height.#2pt}}}
\def\square{\mathord{\dalemb{4.9}{5}\hbox{\hskip1pt}}}

\font\fourteentt=cmtt10 scaled \magstep2
\font\fourteenbf=cmbx12 scaled \magstep1
\font\fourteenrm=cmr12 scaled \magstep1
\font\fourteeni=cmmi12 scaled \magstep1
\font\fourteenssr=cmss12 scaled \magstep1
\font\fourteenmbi=cmmib10 scaled \magstep2
\font\fourteensy=cmsy10 scaled \magstep2
\font\fourteensl=cmsl12 scaled \magstep1
\font\fourteenex=cmex10 scaled \magstep2
\font\fourteenit=cmti12 scaled \magstep1
\font\twelvett=cmtt12 \font\twelvebf=cmbx12
\font\twelverm=cmr12  \font\twelvei=cmmi12
\font\twelvessr=cmss12 \font\twelvembi=cmmib10 scaled \magstep1
\font\twelvesy=cmsy10 scaled \magstep1
\font\twelvesl=cmsl12 \font\twelveex=cmex10 scaled \magstep1
\font\twelveit=cmti12
\font\tenssr=cmss10 \font\tenmbi=cmmib10
\font\nineit=cmti9 
 \font\ninebf=cmbx9
\font\ninerm=cmr9  \font\ninei=cmmi9
\font\ninesy=cmsy9 \font\ninessr=cmss9
\font\ninembi=cmmib10 scaled 900
\font\eightit=cmti8 \font\eightsl=cmsl8
\font\eighttt=cmtt8 \font\eightbf=cmbx8
\font\eightrm=cmr8  \font\eighti=cmmi8
\font\eightsy=cmsy8 \font\eightex=cmex10 scaled 800
\font\eightssr=cmss8 \font\eightmbi=cmmib10 scaled 800
 
\font\sevenbf=cmbx7 \font\sevenrm=cmr7 \font\seveni=cmmi7
\font\sevensy=cmsy7 
\font\sevenssr=cmss9 scaled 778 \font\sevenmbi=cmmib10 scaled 700
 
 \font\sixbf=cmbx7 scaled 875
\font\sixrm=cmr6  \font\sixi=cmmi6
\font\sixsy=cmsy6 \font\sixssr=cmss8 scaled 750
\font\sixmbi=cmmib10 scaled 600
\font\fivessr=cmss8 scaled 625  \font\fivembi=cmmib10 scaled 500

\newskip\ttglue
\newfam\ssrfam
\newfam\mbifam

\mathchardef\alpha="710B
\mathchardef\beta="710C
\mathchardef\gamma="710D
\mathchardef\delta="710E
\mathchardef\epsilon="710F
\mathchardef\zeta="7110
\mathchardef\eta="7111
\mathchardef\theta="7112
\mathchardef\iota="7113
\mathchardef\kappa="7114
\mathchardef\lambda="7115
\mathchardef\mu="7116
\mathchardef\nu="7117
\mathchardef\xi="7118
\mathchardef\pi="7119
\mathchardef\rho="711A
\mathchardef\sigma="711B
\mathchardef\tau="711C
\mathchardef\upsilon="711D
\mathchardef\phi="711E
\mathchardef\chi="711F
\mathchardef\psi="7120
\mathchardef\omega="7121
\mathchardef\varepsilon="7122
\mathchardef\vartheta="7123
\mathchardef\varpi="7124
\mathchardef\varrho="7125
\mathchardef\varsigma="7126
\mathchardef\varphi="7127
\mathchardef\partial="7140

\def\fourteenpoint{\def\rm{\fam0\fourteenrm}
\textfont0=\fourteenrm \scriptfont0=\tenrm \scriptscriptfont0=\sevenrm
\textfont1=\fourteeni \scriptfont1=\teni \scriptscriptfont1=\seveni
\textfont2=\fourteensy \scriptfont2=\tensy \scriptscriptfont2=\sevensy
\textfont3=\fourteenex \scriptfont3=\fourteenex \scriptscriptfont3=\fourteenex
\def\it{\fam\itfam\fourteenit} \textfont\itfam=\fourteenit
\def\sl{\fam\slfam\fourteensl} \textfont\slfam=\fourteensl
\def\bf{\fam\bffam\fourteenbf} \textfont\bffam=\fourteenbf
\scriptfont\bffam=\tenbf \scriptscriptfont\bffam=\sevenbf
\def\tt{\fam\ttfam\fourteentt} \textfont\ttfam=\fourteentt
\def\ssr{\fam\ssrfam\fourteenssr} \textfont\ssrfam=\fourteenssr
\scriptfont\ssrfam=\tenmbi \scriptscriptfont\ssrfam=\sevenmbi
\def\mbi{\fam\mbifam\fourteenmbi} \textfont\mbifam=\fourteenmbi
\scriptfont\mbifam=\tenmbi \scriptscriptfont\mbifam=\sevenmbi
\tt \ttglue=.5em plus .25em minus .15em
\normalbaselineskip=16pt
\bigskipamount=16pt plus5pt minus5pt
\medskipamount=8pt plus3pt minus3pt
\smallskipamount=4pt plus1pt minus1pt
\abovedisplayskip=16pt plus 4pt minus 12pt
\belowdisplayskip=16pt plus 4pt minus 12pt
\abovedisplayshortskip=0pt plus 4pt
\belowdisplayshortskip=9pt plus 4pt minus 6pt
\parskip=5pt plus 1.5pt
\twelvefoot
\setbox\strutbox=\hbox{\vrule height12pt depth5pt width0pt}
\let\sc=\tenrm
\let\big=\fourteenbig \normalbaselines\rm}
\def\fourteenbig#1{{\hbox{$\left#1\vbox to12pt{}\right.\n@space$}}
\def\square{\mathord{\dalemb{6.8}{7}\hbox{\hskip1pt}}}}

\def\twelvepoint{\def\rm{\fam0\twelverm}
\textfont0=\twelverm \scriptfont0=\ninerm \scriptscriptfont0=\sevenrm
\textfont1=\twelvei \scriptfont1=\ninei \scriptscriptfont1=\seveni
\textfont2=\twelvesy \scriptfont2=\ninesy \scriptscriptfont2=\sevensy
\textfont3=\twelveex \scriptfont3=\twelveex \scriptscriptfont3=\twelveex
\def\it{\fam\itfam\twelveit} \textfont\itfam=\twelveit
\def\sl{\fam\slfam\twelvesl} \textfont\slfam=\twelvesl
\def\bf{\fam\bffam\twelvebf} \textfont\bffam=\twelvebf
\scriptfont\bffam=\ninebf \scriptscriptfont\bffam=\sevenbf
\def\tt{\fam\ttfam\twelvett} \textfont\ttfam=\twelvett
\def\ssr{\fam\ssrfam\twelvessr} \textfont\ssrfam=\twelvessr
\scriptfont\ssrfam=\ninessr \scriptscriptfont\ssrfam=\sevenssr
\def\mbi{\fam\mbifam\twelvembi} \textfont\mbifam=\twelvembi
\scriptfont\mbifam=\ninembi \scriptscriptfont\mbifam=\sevenmbi
\tt \ttglue=.5em plus .25em minus .15em
\normalbaselineskip=14pt
\bigskipamount=14pt plus4pt minus4pt
\medskipamount=7pt plus2pt minus2pt
\abovedisplayskip=14pt plus 3pt minus 10pt
\belowdisplayskip=14pt plus 3pt minus 10pt
\abovedisplayshortskip=0pt plus 3pt
\belowdisplayshortskip=8pt plus 3pt minus 5pt
\parskip=3pt plus 1.5pt
\tenfoot
\setbox\strutbox=\hbox{\vrule height10pt depth4pt width0pt}
\let\sc=\ninerm
\let\big=\twelvebig \normalbaselines\rm}
\def\twelvebig#1{{\hbox{$\left#1\vbox to10pt{}\right.\n@space$}}
\def\square{\mathord{\dalemb{5.9}{6}\hbox{\hskip1pt}}}}

\def\tenpoint{\def\rm{\fam0\tenrm}
\textfont0=\tenrm \scriptfont0=\sevenrm \scriptscriptfont0=\fiverm
\textfont1=\teni \scriptfont1=\seveni \scriptscriptfont1=\fivei
\textfont2=\tensy \scriptfont2=\sevensy \scriptscriptfont2=\fivesy
\textfont3=\tenex \scriptfont3=\tenex \scriptscriptfont3=\tenex
\def\it{\fam\itfam\tenit} \textfont\itfam=\tenit
\def\sl{\fam\slfam\tensl} \textfont\slfam=\tensl
\def\bf{\fam\bffam\tenbf} \textfont\bffam=\tenbf
\scriptfont\bffam=\sevenbf \scriptscriptfont\bffam=\fivebf
\def\tt{\fam\ttfam\tentt} \textfont\ttfam=\tentt
\def\ssr{\fam\ssrfam\tenssr} \textfont\ssrfam=\tenssr
\scriptfont\ssrfam=\sevenssr \scriptscriptfont\ssrfam=\fivessr
\def\mbi{\fam\mbifam\tenmbi} \textfont\mbifam=\tenmbi
\scriptfont\mbifam=\sevenmbi \scriptscriptfont\mbifam=\fivembi
\tt \ttglue=.5em plus .25em minus .15em
\normalbaselineskip=12pt
\bigskipamount=12pt plus4pt minus4pt
\medskipamount=6pt plus2pt minus2pt
\abovedisplayskip=12pt plus 3pt minus 9pt
\belowdisplayskip=12pt plus 3pt minus 9pt
\abovedisplayshortskip=0pt plus 3pt
\belowdisplayshortskip=7pt plus 3pt minus 4pt
\parskip=0.0pt plus 1.0pt
\eightfoot
\setbox\strutbox=\hbox{\vrule height8.5pt depth3.5pt width0pt}
\let\sc=\eightrm
\let\big=\tenbig \normalbaselines\rm}
\def\tenbig#1{{\hbox{$\left#1\vbox to8.5pt{}\right.\n@space$}}
\def\square{\mathord{\dalemb{4.9}{5}\hbox{\hskip1pt}}}}

\def\eightpoint{\def\rm{\fam0\eightrm}
\textfont0=\eightrm \scriptfont0=\sixrm \scriptscriptfont0=\fiverm
\textfont1=\eighti \scriptfont1=\sixi \scriptscriptfont1=\fivei
\textfont2=\eightsy \scriptfont2=\sixsy \scriptscriptfont2=\fivesy
\textfont3=\eightex \scriptfont3=\eightex \scriptscriptfont3=\eightex
\def\it{\fam\itfam\eightit} \textfont\itfam=\eightit
\def\sl{\fam\slfam\eightsl} \textfont\slfam=\eightsl
\def\bf{\fam\bffam\eightbf} \textfont\bffam=\eightbf
\scriptfont\bffam=\sixbf \scriptscriptfont\bffam=\fivebf
\def\tt{\fam\ttfam\eighttt} \textfont\ttfam=\eighttt
\def\ssr{\fam\ssrfam\eightssr} \textfont\ssrfam=\eightssr
\scriptfont\ssrfam=\sixssr \scriptscriptfont\ssrfam=\fivessr
\def\mbi{\fam\mbifam\eightmbi} \textfont\mbifam=\eightmbi
\scriptfont\mbifam=\sixmbi \scriptscriptfont\mbifam=\fivembi
\tt \ttglue=.5em plus .25em minus .15em
\normalbaselineskip=9pt
\bigskipamount=9pt plus3pt minus3pt
\medskipamount=5pt plus2pt minus2pt
\abovedisplayskip=9pt plus 3pt minus 9pt
\belowdisplayskip=9pt plus 3pt minus 9pt
\abovedisplayshortskip=0pt plus 3pt
\belowdisplayshortskip=5pt plus 3pt minus 4pt
\parskip=0.0pt plus 1.0pt
\setbox\strutbox=\hbox{\vrule height8.5pt depth3.5pt width0pt}
\let\sc=\sixrm
\let\big=\eightbig \normalbaselines\rm}
\def\eightbig#1{{\hbox{$\left#1\vbox to6.5pt{}\right.\n@space$}}
\def\square{\mathord{\dalemb{3.9}{4}\hbox{\hskip1pt}}}}

\def\vfootnote#1{\insert\footins\bgroup\footsuite
    \interlinepenalty=\interfootnotelinepenalty
    \splittopskip=\ht\strutbox
    \splitmaxdepth=\dp\strutbox \floatingpenalty=20000
    \leftskip=0pt \rightskip=0pt \spaceskip=0pt \xspaceskip=0pt
    \textindent{#1}\footstrut\futurelet\next\fo@t}
\def\hangfootnote#1{\edef\@sf{\spacefactor\the\spacefactor}#1\@sf
    \insert\footins\bgroup\footsuite
    \let\par=\endgraf
    \interlinepenalty=\interfootnotelinepenalty
    \splittopskip=\ht\strutbox
    \splitmaxdepth=\dp\strutbox \floatingpenalty=20000
    \leftskip=0pt \rightskip=0pt \spaceskip=0pt \xspaceskip=0pt
    \smallskip\item{#1}\bgroup\strut\aftergroup\@foot\let\next}
\def\footsuite{}
\def\twelvefoot{\def\footsuite{\twelvepoint}}
\def\tenfoot{\def\footsuite{\tenpoint}}
\def\eightfoot{\def\footsuite{\eightpoint}}
\catcode`@=12

\def\semidirprod{\rlap{\ssr C}\raise1pt\hbox{$\mkern.75mu\times$}}
\def\i{{\rm i}}
\def\x{\hbox{\ssr x}}
\def\ds{\diamondsuit}
\def\px{\phantom{\x}}
\def\for{\lower6pt\hbox{$\Big|$}}
\def\R{{\rm\rlap I\mkern3mu R}}
\vsize=9in
\twelvepoint\nofirstpagenotwelve
\oneandahalfspace
\rightline{IC/92/122}
\rightline{Imperial/TP/90-91/21}
\rightline{CTP TAMU-26/91}
\rightline{June, 1992}
\vskip 2cm\fourteenpoint
\centerline{Nonlinear Realisations of $w_{1+\infty}$}
\twelvepoint
\bigskip\bigskip
\centerline{E. Sezgin}
\medskip
\centerline{\it Center for Theoretical Physics}
\centerline{\it Texas A\&M University}
\centerline{\it College Station, TX 77843, U.S.A.}
\bigskip
\centerline{and}
\bigskip
\centerline{K.S. Stelle}
\medskip
\centerline{\it The Blackett Laboratory}
\centerline{\it Imperial College}
\centerline{\it Prince Consort Road}
\centerline{\it London SW7 2BZ, U.K.}
\bigskip
\AB\singlespace
The nonlinear scalar-field realisation of $w_{1+\infty}$ symmetry in $d=2$
dimensions is studied in analogy to the nonlinear realisation of $d=4$
conformal symmetry $SO(4,2)$. The $w_{1+\infty}$ realisation is derived
from a coset-space construction in which the divisor group is generated by
the non-negative modes of the Virasoro algebra, with subsequent
application of an infinite set of covariant constraints.  The initial
doubly-infinite set of Goldstone fields arising in this construction is
reduced by the covariant constraints to a singly-infinite set
corresponding to the Cartan-subalgebra generators $v^\ell_{-(\ell+1)}$.
We derive the transformation rules of this surviving set of fields,
finding a triangular structure in which fields transform into themselves
or into lower members of the set only. This triangular structure gives
rise to finite-component subrealisations, including the standard one for a
single scalar.  We derive the Maurer-Cartan form and discuss the
construction of invariant actions.
\AE\oneandahalfspace
\np
\noindent{\bf 1. Introduction}
\bigskip
The $w_{1+\infty}$ algebra that has been studied as a
higher-spin symmetry algebra in $d=2$ conformal field theories admits its
simplest field-theoretic realisation in terms of a single scalar field
$\varphi(x)$.  This realisation is necessarily nonlinear because its
field content is far too small to support a linear realisation of the
algebra. $w_{1+\infty}$ can naturally be viewed as the algebra of
symplectic diffeormphisms of a two-dimensional cylinder $(y,\theta)$;
this linear realisation can in turn be viewed as a Poisson bracket
algebra of the basis functions on the cylinder
$$
{\rm v}^\ell_m=-\i y^{\ell+1} e^{\i m\theta},\eqno(1)
$$
with a resulting algebra of differential operators
$v^\ell_m=e^{\i m\theta}(my^{\ell+1}\partial/\partial
y+\i(\ell+1)y^\ell\partial/\partial\theta)$ given by
$$
[v^j_m,
v^\ell_n] = [(\ell+1)m-(j+1)n] v^{j+\ell}_{m+n}\qquad
\ell, j\ge-1,\ -\infty<m, n<\infty.\eqno(2)
$$
For the $w_\infty$ algebra, the upper indices are restricted to values
$\ell\ge0$.  Both $w_{1+\infty}$ and $w_\infty$ contain the Virasoro algebra
generated by $v^0_m$.

Since only one of the $d=2$ worldsheet coordinates of $\varphi(x)$ is
involved in the $w_\infty$ transformations, this field is effectively a
function of only one variable in as far as the realisation is concerned,
and is thus insufficient to support a full linear realisation of the algebra
(2). The
nonlinear realisation on $\varphi$,
$$
\delta\varphi = k_\ell(\partial_+\varphi)^{\ell+1},\eqno(3)
$$
with parameters $k_\ell(x^+)$ that are ``semilocal'', since they depend
on $x^+$ but not $x^-$, may be viewed as arising from a coset-space
construction $w_{1+\infty}/w_{\infty}$ [1] in which the coset parameter
may be considered to be a function of just the $\theta$ variable on the
cylinder, since the basis functions in (1) that are left out of
$w_\infty$ are independent of $y$.  The field $\varphi$ may be identified
with this coset parameter provided $x^+$ is identified with $\theta$.
Note that $x^-$ here is a ``time'' unrelated to this chiral
algebra.

The derivation of the transformations (3) from the
$w_{1+\infty}/w_{\infty}$ coset construction given in [1]
follows standard techniques of the theory of nonlinear realisations
[2, 3].  An unusual feature of this construction, however, is the fact
that the transformations of the Goldstone field $\varphi(x)$ are
nonlinear also for the {\it divisor} group $w_\infty$, and not just for
the transformations belonging to the coset.  This is due to the fact that
the $w_{1+\infty}/w_{\infty}$ coset is {\it non-reductive,} {\it
i.e.}\ the coset generators do not form a linear representation of the
divisor group $w_\infty$, as can be seen in (2), since commutators of
coset generators with $w_\infty$ generators produce results lying
mostly in $w_\infty$ and not in the coset.  The only generators in
$w_\infty$ with respect to which $\varphi(x)$ actually transforms linearly
are the Virasoro generators $v^0_m$, as can be seen in (3).

In this paper, we seek a more detailed understanding of the
group-theoretic aspects of the nonlinear realisation (3) by starting from an
essentially
reductive coset construction in which we choose the divisor group to be
generated by the $m\ge0$ modes of the Virasoro algebra, which we shall
denote by ${\rm Vir}^+$. In this construction, we shall view the Virasoro
algebra as a conformal algebra on the circle $S^1$, working in a chart
where the $x^+$ coordinate is itself viewed as a coset parameter
associated to the Virasoro generator $L_{-1}=v^0_{-1}$.  In this
realisation, the generators correspond to a Laurent expansion of the
basis functions instead of the Fourier expansion used in (1).  The
Virasoro generators in this basis, which act only on the $x^+$ dependence
of $\varphi(x)$ since it is a scalar, are then the differential operators
$$
v^0_{m}=-(x^+)^{m+1}{\partial\over\partial x^+}.\eqno(4)
$$
The nonsingular generators are those for which $m\ge-1$; the
corresponding nonsingular Virasoro subalgebra shall be denoted ${\rm
Vir}^\uparrow$.  This may be extended to a nonsingular subalgebra of
$w_{1+\infty}$, denoted by $w^\uparrow_{1+\infty}$, that is generated by
$v^\ell_m$ with $m\ge-\ell-1$. Our coset construction will be based on
$w^\uparrow_{1+\infty}/{\rm Vir}^+$.   We shall consider
this coset to be ``essentially reductive'' in the sense that the divisor
subgroup transformations of all Goldstone {\it fields} are linear ({\it
i.e.}\  all the coset generators except $v^0_{-1}\leftrightarrow x^+$ form a
linear realisation of the divisor subgroup ${\rm Vir}^+$).
\bigskip
\noindent{\bf 2. The $SO(4,2)/SO(3,1)$ analogy} \bigskip In order to set the
stage for our later discussion of $w_{1+\infty}$, we recall first a more
familiar non-reductive coset construction: the Minkowski-space realisation
of $d=4$ conformal symmetry. The group $SO(4,2)$ contains two Poincar\'e
subgroups --- the usual one, which we shall  denote by $P$, composed of
$SO(3,1)\semidirprod\{P^\mu\}$, {\it i.e.} the Lorentz subgroup taken in a
semidirect product with the translations $P^\mu$, and also an unorthodox
one, $P'$, in which the ``translational'' generators are the proper
conformal generators $K^\mu$. Minkowski space may be realised as the coset
$SO(4,2)/(P'\times\{D\})$, leaving the $P^\mu$ generators in the coset [4].
Owing to the fact that the subgroup $P'\times\{D\}$ is a maximal subgroup of
$SO(4,2)$, this coset space is actually compact, and in fact has the global
topology $S^3 \times S^1$. Accordingly, it is known as compactified
Minkowski space ${\cal M}^\sharp$.

The $SO(4,2)/(P'\times\{D\})$ coset construction yields nonlinear
transformations of the coset parameter $x^\mu$ in the usual fashion via
multiplication on the left by an arbitrary element of the group and then
factorisation of the result into (coset element)$\times$(divisor group
element).  In this fashion, one recovers the usual Poincar\'e group
transformations together with the proper conformal transformations:
$$
x^\mu\to x'^\mu = {x^\mu - c^\mu x^2\over 1 - 2c\cdot x + c^2x^2}.\eqno(5)
$$
This coset construction is nonreductive since $[K_\mu,P_\nu] =
-2\i(\eta_{\mu\nu}D + M_{\mu\nu})$, so the coset generators $P_\nu$ do not
form a linear representation of the divisor group.  As a result, the proper
conformal transformations of $x^\mu$ generated by the $K_\mu$ are
nonlinear even though they belong to the divisor group of our coset.

The conformal transformations of Minkowski space may alternatively be
obtained in a way that makes use of a reductive coset-space
construction.  This will necessarily involve a larger coset
than the
above non-reductive construction.  Since we want to maintain Lorentz
covariance, the divisor group must contain the Lorentz group.  Accordingly,
we consider the coset $SO(4,2)/SO(3,1)$, which requires 9 coset
parameters.  Following Ref.\ [3] for non-linear realisations of
spacetime symmetries, we let the 9 coset parameters be represented by the
four $x^\mu\leftrightarrow P^\mu$ and by five Goldstone fields, which are
taken to be functions of $x^\mu$:  $\varphi(x)\leftrightarrow D$ and
$b^\mu(x)\leftrightarrow K^\mu$.

The 9 coset parameters $x^\mu$, $\varphi(x)$ and $b^\mu(x)$
evidently form linear representations of the divisor group $H=SO(3,1)$.
They also form a larger coset than we had in the non-reductive construction
above.  In this case, however, not all of the coset parameters are really
essential, for we may subsequently eliminate the proper conformal Goldstone
fields by covariant constraints; this procedure has been called the
``inverse Higgs effect'' [5, 6].

In deriving the covariant constraints of the inverse Higgs effect, it is
appropriate to use Maurer-Cartan forms, starting from a coset element
written using the standard exponential parametrisation,
$$
k = e^{\i x^\mu P_\mu} e^{\i \varphi(x) D} e^{\i b^\mu(x) K_\mu}.\eqno(6)
$$
The Maurer-Cartan forms are then given by decomposing the Lie algebra
element $k^{-1}dk$ into its various projections in the Lie algebra:
$${\cal P} = k^{-1}dk = \i \omega^\mu_PP_\mu + \i \omega^\mu_KK_\mu
+ \i \omega_D D + \i
\omega_{H\,\rho\sigma}M^{\rho\sigma}.\eqno(7)
$$

The Maurer-Cartan forms $\omega^\mu_P$, $\omega^\mu_K$ and $\omega^\mu_D$
belonging to the coset transform in a standard way according to their $H
= SO(3,1)$ indices, but with {\it field-dependent parameters} when
transformed by group elements in the coset.  The parameters for these
field-dependent $H$ transformations are found by left multiplication and
repolarisation into a product $k'h'$, where $h'\in H$:
$$
g_0k = k'(g_0;x^\mu,\varphi(x),b^\mu(x))\,h'(g_0;x^\mu,\varphi(x),
b^\mu(x));\eqno(8)
$$
the transformed values of the Goldstone fields and
$x^\mu$ are then given by rewriting $k'$ in the form (6),
$$
k'=e^{\i x'^\mu P_\mu}e^{\i\varphi'(x')D}e^{\i b'^\mu(x')K_\mu}.\eqno(9)
$$
The Maurer-Cartan forms can be calculated from (7). In particular,
one finds
that
$$
\omega_\mu^P = \partial_\mu \varphi -2 b_\mu. \eqno(10)
$$
which shows that the independent Goldstone field $b_\mu$ is inessential
since we can covariantly impose the constraint
$$
\omega_\mu^P =0 \Rightarrow b_\mu={1\over 2}\partial_\mu \varphi.
\eqno(11)
$$

The possibility of eliminating the independent $b_\mu$ Goldstone field here
stems
in part from the fact that there are two cosets that one could take for a
nonlinear realisation involving a dilaton Goldstone field:  $SO(4,2)/P'$ and
$SO(4,2)/SO(3,1)$.  In other words, there is a choice as to whether the
$K_\mu$
generators belong to the coset or are divided out by being included into the
divisor.  The constraint (11) returns us to the field content of the
nonlinear realisation based on the smaller coset $SO(4,2)/P'$, after
having initially started out with the larger coset $SO(4,2)/SO(3,1)$.

Having found, {\it via} the inverse Higgs constraint (11), that one may
construct realisations of the $SO(4,2)$ symmetry with only a dilaton
Goldstone field $\varphi(x)$, one may then ask under what conditions
the full $SO(4,2)$ symmetry may be realised without even this Goldstone
field, {\it i.e.}\ on the minimal coset, $SO(4,2)/(P'\times\{D\})={\cal
M}^\sharp$.  Such a realisation is indeed possible, but since there are no
further covariant constraints to impose, realising $SO(4,2)$ symmetry on
${\cal M}^\sharp$ requires a new feature: {\it gauge invariance} of the
action with respect to local $D$ transformations.  A well-known example of
this situation occurs in classical Yang-Mills theory, which can be
formulated in terms of a $SO(4,2)/P'$ realisation, but it then turns out
that the $\varphi(x)$ Goldstone field for the $D$ transformations
decouples, or ``drops out'', of the action --- {\it i.e.} there is a local
$D$ invariance for the system comprising both the Yang-Mills fields and the
dilaton field $\varphi(x)$.  Consequently, one may view Yang-Mills
theory as a realisation of $SO(4,2)$ on its smallest coset space
$SO(4,2)/(P'\times\{D\})$. Such occurrences are clearly ``accidental''
from the standpoint of nonlinear realisation theory --- there is no way in
which one could adjust the Yang-Mills Lagrangian for the vector gauge
field alone in order to achieve this local $D$ invariance had it not been
present.  Indeed, at the quantum level, this local symmetry is generally lost
as a result of the familiar trace anomaly.  A standard way of calculating
this anomaly in fact is to introduce a dilaton field $\varphi(x)$ and
to compute its purely quantum-induced coupling.

This discussion of the $SO(4,2)$ realisations in $d=4$ will be instructive
for our purposes because many of its features have direct analogues
in the $w_{1+\infty}$ realisations to which we shall now turn.  As
mentioned above and as discussed in Ref.\ [1], the minimal
$w_{1+\infty}/w_\infty$ coset space that we can use to realise $w_{1+\infty}$
symmetry on scalar fields is also non-reductive.   We shall see below that we
may reformulate this realisation using an essentially reductive coset space
construction followed by an imposition of covariant constraints.
Finally, we shall see that the single-scalar realisation of $w_{1+\infty}$
is obtained thanks to the possibility of requiring extra gauge symmetries
analogous to the local $D$ symmetry discussed above.
\bigskip
\noindent{\bf 3. $w_{1+\infty}^\uparrow/{\rm Vir}^+$}
\bigskip
Let us now return to the
$w_{1+\infty}^\uparrow$ algebra given in Eq.\ (1).  We shall
concentrate on realisations of this
algebra and of its associated group instead of the full $w_{1+\infty}$
because in this way
we may pursue more closely the $d=4$ conformal analogy given in section 2.
The realisation of $w_{1+\infty}^\uparrow$ given here generalises the
realisation of Virasoro symmetry on a single coordinate $x^+$ used in [7,
8], in which $x^+$ is interpreted as the coordinate of a coset ${\rm
Vir}^\uparrow/{\rm Vir}^+$, {\it i.e.}\ as the coset parameter
corresponding to $L_{-1}=v^0_{-1}$.  Since the ${\rm
Vir}^\uparrow/{\rm Vir}^+$ coset is itself non-reductive, the
divisor-group ${\rm Vir}^+$ transformations of $x^+$ are
nonlinear, in analogy to (5):
$$
\delta x^+=k_n(x^+)^{n+1}.\eqno(12)
$$
The restriction to ${\rm Vir}^\uparrow$ corresponds to the
non-singular Virasoro generators in this realisation, and
correspondingly for $w_{1+\infty}^\uparrow$.

Consider now the Maurer-Cartan decomposition for the coset
$w^\uparrow_{1+\infty}/{\rm Vir}^+$:
$$
   v^i_m \quad \rightarrow \quad v^0_m \ (m\ge 0)\ \oplus\ \{v^0_{-1},\
v^\ell_m ;\
 \ \ \ell\ne 0, \ m\ge -\ell-1\},  \eqno(13)
$$
where the $v^0_{m\ge0}$ are the ${\rm Vir}^+$ generators.  In analogy with
the case of $SO(4,2)$ discussed earlier, we choose the coset
representative
$$
    k=e^{-x^+ v^0_{-1}}\prod_{\ell\ne 0}e^{-\phi^{\{\ell\}}}, \eqno(14)
$$
where we have used the notation
$$
    \phi^{\{\ell\}} \equiv \sum_{m=-\ell-1}^\infty \phi^\ell_m v^\ell_m.
\eqno(15)
$$
Note that we have now chosen to work with antihermitean generators.
{}From (15) one sees that there are infinitely many initial Goldstone fields
$\phi^\ell_m$ corresponding to the coset generators $v^\ell_m,\  \ell\ne 0$.

Next we consider the Maurer-Cartan form, which reads
$$
{\cal P}=k^{-1}dk= \sum_{m\ge 0}E^{-1}_mv^{-1}_m +E^0_{-1}v^0_{-1}+
\sum_{\ell\ge 1 \atop m\ge -\ell-1}E^\ell_mv^\ell_m + \sum_{m\ge
0}\omega^0_m v^0_m,  \eqno(16)
$$
where $E^\ell_m$ and $\omega^0_m$ are all 1-forms, {\it i.e.}\
$E^\ell_m=dx^+E_{(+)\, m}^\ell+dx^-E_{(-)\, m}^\ell$, {\it etc.}
The ``vierbein'' components $E_{(\pm)m}^{-1}$ and $E_{(\pm)m}^{\ell\ge1}$
belong to linear representations of the divisor group ${\rm Vir}^+$, and
hence will transform homogeneously under $w^\uparrow_{1+\infty}$
transformations, albeit with field-dependent parameters when the
transformations are taken from $w^\uparrow_\infty/{\rm
Vir}^\uparrow$.  This homogeneous transformation property is the benefit
that we derive from the essentially reductive structure of the coset
$w_{1+\infty}^\uparrow/{\rm Vir}^+$.

In the computation of the various components of ${\cal P}$,
the following formulas [8] will prove useful:
$$
e^\phi \beta e^{-\phi}=e^\phi \wedge \beta,\quad \quad e^\phi d e^{-\phi}=
\bigg({1-e^\phi\over \phi}\bigg)\wedge d\phi,  \eqno(17)
$$
where we have used the notation
$$
   \phi\wedge \beta \equiv [\phi,\beta],\qquad\qquad
\phi^2\wedge\beta\equiv\phi\wedge\phi\wedge\beta=[\phi,[\phi,\beta]],\
{\it etc.}
\eqno(18)
$$
Note that the wedge $(\wedge)$ notation used here denotes an operation
involving
multiple $\omega_{1+\infty}$ commutators, and is not to be confused with the
exterior product for forms.  All equations containing forms in this paper will
involve 1-forms without exterior products.  Using (17, 18) to evaluate
(16), we find
$$
\eqalign{
&\cdots e^{\phi^{\{4\}}}\cdots
e^{\phi^{\{1\}}}\wedge\big(-dx^+v^0_{-1}-\phi^{\{-1\}}\wedge dx^+
-d\phi^{\{-1\}}\big)\cr
&\phantom{\cdots}+\cdots e^{\phi^{\{4\}}}\cdots e^{\phi^{\{2\}}}
\bigg({1-e^{\phi^{\{1\}}}\over \phi^{\{1\}}}\bigg)\wedge d\phi^{\{1\}} +\cdots
e^{\phi^{\{4\}}} e^{\phi^{\{3\}}} \bigg({1-e^{\phi^{\{2\}}}\over
\phi^{\{2\}}}\bigg)\wedge d\phi^{\{2\}} +\cdots\cr
&\phantom{\cdots
e^{\phi^{\{4\}}}\cdots e^{\phi^{\{1\}}}\big(-dx^+} = \sum_{m\ge 0}E^{-
1}_mv^{-1}_m + E^0_{-1}v^0_{-1}+\sum_{\ell\ge 1 \atop m\ge
-\ell-1}E^\ell_mv^\ell_m +\sum_{m\ge 0}\omega^0_m v^0_m.\cr} \eqno(19)
$$

      We can now compute the forms $E^\ell_m$ and $\omega^0_m$ by equating
terms proportional to $v^\ell_m$ and $v^0_m$, respectively. We shall refer
to the spin label $\ell$ as the level of a generator.  From the $(+)$
component of (19) at level $\ell=-1$, we read off the equation
$$
  \sum_{m\ge 0}E_{(+)\,m}^{-1}v^{-1}_m=
-\partial_+\phi^{\{-1\}}-\phi^{\{-1\}}\wedge (v^0_{-1}).
\eqno(20)
$$
We now observe that one can impose the covariant
constraint $E_{(+)\,m}^{-1}=0$. This constraint enables us to solve
algebraically for $\phi^{-1}_{m\ge1}$ in terms of $\partial_+$ derivatives of
$\phi^{-1}_0$.  Specifically, we have
$$
E_{(+)\, m}^{-1}=0\ \ \Rightarrow\ \
\partial_+\phi^{-1}_m+(m+1)\phi^{-1}_{m+1}=0, \ \ m\ge 0.\eqno(21)
$$
Note that $\phi^{-1}_0$ is the only $\ell=-1$
Goldstone field that is not eliminated as an independent field by this
constraint.  Using the $\ell=-1$ constraint (21), we find at level $\ell=0$
$$\eqalignno{
E_{(+)\,-1}^0&=-dx^+&(22)\cr
\omega_{(+)\,m}^0&=0.&(23)\cr}
$$
Thus, the constraint (21) has the consequence that all of the
connection terms $\omega_{(+)\,m}^0$ vanish.

     The remainder of the $(+)$ component of the form $\cal P$ at levels
$\ell\ge 1$ transforms homogeneously and thus can also be set to zero.
This gives us the  maximum number of constraints with which we can eliminate
inessential Goldstone  fields.  Anticipating the answer, we observe that
substituting
$$
\partial_+\phi^{\{\ell\}} + \phi^{\{\ell\}}\wedge v^0_{-1} =0,\ \ \ \ \ell\ne
0,
\eqno(24)
$$
into the $(+)$ component of ${\cal P}$ gives rise to pairwise cancellation of
all
terms except $-dx^+$. Conversely, we can derive (22--24) from the set of
covariant  constraints
$$
\cases{
E_{(+)\,m}^{-1} = 0, \ \ m\ge 0\cr
E_{(+)\,m}^\ell = 0, \ \ \ell\ge 1,\ \ m\ge -\ell-1.\cr}\eqno(25)
$$

{}From (19), we can also compute $E_{(-)m}^\ell$ and $\omega_{(-)m}^0$. The
first few levels of the $E_{(-)m}^\ell$ are given by
$$
\eqalignno{
 E_{(-)m}^{-1}&= -\partial_-\phi^{-1}_m,  &(26)\cr
 E_{(-)-1}^{0}&=2\phi^1_{-2}\partial_-\phi^{-1}_1,  &(27)\cr}
$$
while for the $\omega_{(-)m}^0$ the result is
$$
\omega_{(-)m}^0=2\sum_{n\ge 1}
n\phi^1_{m-n}\partial_-\phi^{-1}_n.\eqno(28)
$$

We may summarise the results of our inverse-Higgs-effect analysis by the
following diagram of the $w^\uparrow_{1+\infty}$ generators:
\bigskip
\leftline{Fig.\ 1\hskip 5cm\underscore{The generators of
$w^\uparrow_{1+\infty}$}}
\nobreak
$$
\matrix{\px&\vdots&\vdots&\vdots&\vdots&\vdots&\vdots&\vdots&
\vdots&\vdots&\vdots&\vdots&\vdots&\vdots&\vdots&\vdots&\vdots&\cr
  \px&\circ&\x&\x&\x&\x&\x&\x&\x&\x&\x&\x&\x&\x&\x&\x&\x&\cdots\cr
    \px&&\circ&\x&\x&\x&\x&\x&\x&\x&\x&\x&\x&\x&\x&\x&\x&\cdots\cr
      \px&&&\circ&\x&\x&\x&\x&\x&\x&\x&\x&\x&\x&\x&\x&\x&\cdots\cr
        \px&&&&\circ&\x&\x&\x&\x&\x&\x&\x&\x&\x&\x&\x&\x&\cdots\cr
          \px&&&&&\circ&\x&\x&\x&\x&\x&\x&\x&\x&\x&\x&\x&\cdots\cr
            \px&&&&&&\circ&\x&\x&\x&\x&\x&\x&\x&\x&\x&\x&\cdots\cr
   \smash{\vdots}&&&&&&&\circ&\x&\x&\x&\x&\x&\x&\x&\x&\x&\cdots\cr
             \ell=2&&&&&&&&\circ&\x&\x&\x&\x&\x&\x&\x&\x&\cdots\cr
               \ell=1&&&&&&&&&\circ&\x&\x&\x&\x&\x&\x&\x&\cdots\cr
\ell=0&-&-&-&-&-&-&-&-&-&\bullet&\ds&\ds&\ds&\ds&\ds&\ds&\cdots\cr
                  \ell=-1&&&&&&&&&&&\circ&\x&\x&\x&\x&\x&\cdots\cr
                                         &&&&&&&&&&&\vert&&&&&&\cr
                       &&&&&&&&&&\px\rlap{m}&=&\llap{0}\px&&&&&\cr}
$$
In this diagram, the generators corresponding to reducible Goldstone fields
eliminable by covariant constraints in the inverse Higgs effect are
indicated by $\x$, the irreducible Goldstone fields surviving the inverse
Higgs effect are indicated by $\circ$, the $v^0_{-1}$ generator associated
to the coordinate $x^+$ is indicated by $\bullet$, and the generators of the
divisor subalgebra ${\rm Vir}^+$ are indicated by $\ds$.
\bigskip
\noindent{\bf 4. Transformation rules for the Goldstone fields}
\bigskip
We now derive the transformation rules
for the surviving Goldstone fields lying on the left edge of the Fig.\ 1
diagram.  Similarly to the $SO(4,2)$ case given in Eq.\ (8),
the action of the group $w_{1+\infty}^\uparrow$ on a coset
representative, which we shall generically denote by $e^{-\phi(x)}$, is as
follows:
$$
        g e^{-\phi(x)} =e^{-\phi'(x')} h,  \eqno(29)
$$
where $h$ is an element of the divisor subgroup ${\rm Vir}^\uparrow$. For
infinitesimal transformations, we have
$$
        e^{\phi(x)+{\bar \delta}\phi(x)}(1+\delta g) e^{-\phi(x)}=
1+\delta h, \eqno(30)
$$
where
$$
   \eqalign{
             {\bar \delta}\phi(x) &\equiv \phi'(x')-\phi(x)  \cr
                           & =\phi'(x)+\delta x^+\partial_+\phi(x)-
\phi(x) \cr
     &\equiv \delta\phi(x)+\delta x^+\partial_+\phi(x).\cr}  \eqno(31)
$$
We use in our derivations a version of the theory of nonlinear
realisations adapted specifically to the realisation of spacetime
symmetries [3].  This shall give us the Einstein-style transformation
$\bar\delta\phi$ directly. However, we shall subsequently view
these transformations from an {\it active} viewpoint, in which the
variation of the field is taken to be the quantity $\delta\phi$ as defined in
(31).  Thus, the transformations of $x^+$ that would occur in Einstein-style
transformations generated by the $v^\ell_m$ will be replaced in the active
viewpoint by transport terms.  These transport terms will be
field-dependent for the $v^{\ell\ge1}_m$ but not for the ${\rm Vir}^+$
generators, as one can see in (4).  As defined in the
introduction, this is what we mean by an essentially reductive coset
construction.  Projecting (30) into the coset direction yields the formula
({\it c.f.}\ [9])
$$ \eqalign{
    \Big(e^\phi{\bar \delta}e^{-\phi}\Big)\for_{G/H}
    &=\Big(e^\phi\delta g e^{-\phi}\Big)\for_{G/H}\cr}. \eqno(32)
$$
Upon the use of the constraint (24), the variation (31) simplifies to
$$
{\bar \delta}\phi^{\{\ell\}} = \delta \phi^{\{\ell\}}-\phi^{\{\ell\}} \wedge
\delta x^+v^0_{-1}. \eqno(33)
$$
Substituting this result in (32) we find
$$
  \eqalignno{
          &\cdots e^{\phi^{\{4\}}}\cdots e^{\phi^{\{1\}}}\big(-\delta
x^+v^0_{-1 } -\phi^{\{-1\}}\wedge
         \delta x^+v^0_{-1}\big) -\cdots e^{\phi^{\{4\}}}\cdots
e^{\phi^{\{1\}}} \wedge
            (\delta\phi^{\{-1\}}-\phi^{\{-1\}}\wedge \delta x^+v^0_{-1})\cr
          &\phantom{\cdots e^{\phi^{\{4\}}}} +\cdots e^{\phi^{\{4\}}}\cdots
e^{\phi^{\{2\}}}
            \bigg({1-e^{\phi^{\{1\}}}\over \phi^{\{1\}}}\bigg)\wedge
(\delta\phi^{\{1\}} -\phi^{\{1\}}\wedge \delta x^+v^0_{-1})\cr
&\phantom{\cdots e^{\phi^{\{4\}}}\cdots e^{\phi^{\{1\}}}} +\cdots
e^{\phi^{\{4\}}} e^{\phi^{\{3\}}} \bigg({1-e^{\phi^{\{2\}}}\over
\phi^{\{2\}}}\bigg)\wedge
              (\delta\phi^{\{2\}}-\phi^{\{2\}}\wedge\delta
x^+v^0_{-1})+\cdots\cr &\phantom{\cdots e^{\phi^{\{4\}}}\cdots
e^{\phi^{\{1\}}} \cdots e^{\phi^{\{4\}}}\cdots e^{\phi^{\{1\}}}}= \cdots
e^{\phi^{\{4\}}}e^{\phi^{\{3\}}}e^{\phi^{\{2\}}}
e^{\phi^{\{1\}}}e^{\phi^{\{-1\}}} e^{x^+v^0_{-1}}\wedge \delta g. &(34)\cr}
$$
Note that this expression is obtainable from the $(+)$ component of (19)
by the replacements $dx^+\rightarrow \delta x^+$ and
$d\phi^{\{\ell\}}\rightarrow {\bar \delta}\phi^{\{\ell\}}=\delta\phi
-\phi\wedge\delta x^+v^0_{-1}$. Just as all the $dx^+$ terms except the
$-dx^+$ in the first term in (19) cancel pairwise by virtue of the
constraint (22), so do all the $\delta x^+$ terms cancel pairwise in the
above equation except the $-\delta x^+$ in the first term.  To simplify
Eq.\ (34) further, we calculate the first two levels of ``dressing'' of
$\delta g$, {\it i.e.}\ we calculate $e^{\phi^{\{-1\}}}
e^{x^+v^0_{-1}}\wedge \delta g $. Let us parametrise $\delta g$ as follows $$
\delta g =\sum_{\ell, m} \alpha^\ell_m
v^\ell_m, \eqno(35)
$$
where the $\alpha^\ell_m$ are $x^+$-independent parameters.
Consider a spin-$\ell$ transformation with parameter $\alpha^{\{\ell\}}
\equiv \sum_m \alpha^\ell_m v^\ell_m$. We find that, analogously to (20),
$$
e^{x^+v^0_{-1}}\alpha^{\{\ell\}}e^{-x^+v^0_{-1}}=
e^{x^+v^0_{-1}}\wedge\alpha^{\{\ell\}}=\sum_m \beta^\ell_m (x^+) v^\ell_m,
\eqno(36)
$$ where the dressed $x^+$-dependent parameters $\beta^\ell_{m+1}(x^+)$
are
$$
    \beta^\ell_{m+1}(x^+) = \sum_{p=0}^{\infty}(-1)^p{(\ell+m+2+p)!
\over{(\ell+m+2)!\,p!}} (x^+)^p \alpha^\ell_{m+1+p}.\eqno(37)
$$
It follows that
$$
\beta^\ell_{m+1}(x^+)={-1\over {\ell+m+2}}\partial_+\beta^\ell_m(x^+).
\eqno(38)
$$
Note that this is the same relation as that satisfied by the fields
$\phi^\ell_m$. Proceeding on to the next level of dressing, with
$\phi^{\{-1\}}$, we evaluate
$$
\sum_{\ell m}e^{\phi^{\{-1\}}} \beta^\ell_m v^\ell_m e^{-\phi^{\{-1\}}}
     =\sum_{\ell m} \gamma^\ell_m(\phi^{\{-1\}})v^\ell_m,  \eqno(39)
$$
where
$$
  \gamma^\ell_m = \beta^\ell_m +\sum_{k \{n_k \} } {1\over k!}
(\ell+2)(\ell+3)\cdots(\ell+k+1)
        (n_1\phi^{-1}_{n_1})(n_2\phi^{-1}_{n_2})\cdots (n_k\phi^{-1}_{n_k})
           \beta^{\ell+k}_{m-n_1-n_2-\cdots -n_k}.   \eqno(40)
$$
{}From this expression, we learn that
$$
\gamma^{\ell+1}_{m-n} ={1\over n(\ell+2)}{\delta\gamma^\ell_m\over
\delta\phi^{-1}_n}. \eqno(41)
$$
Hence, for the $\gamma$-parameters corresponding to the left edge of the
Fig.\ 1 diagram, we have the relation
$$
\gamma^\ell_{-\ell-1} ={1\over (\ell+1)!}{\delta^{\ell+1}
\gamma^{-1}_0 \over \delta y^{\ell+1}}, \eqno(42)
$$
where we have introduced the notation
$$
   y\equiv -\partial_+ \phi^{-1}_0.    \eqno(43)
$$

Turning back to Eq.\ (29) and comparing the $\ell=0$ and $\ell=-1$ terms on
both sides, we learn that
$$\eqalignno{
           \delta x^+&=-\gamma^0_{-1}&(44a)\cr
           \delta \phi^{\{-1\}}&=-\gamma^{\{-1\}}.&(44b)\cr}
$$
Then, with the notation
$$
    \beta^\ell_{-\ell-1}(x^+)\equiv k^\ell(x^+) \eqno(45)
$$
and $y$ as defined in (43), we find that
$$
\delta x^+=\sum_{\ell=0}^\infty(\ell+1) k^\ell y^\ell\eqno(46)
$$
and that the transformation rule (44$b$) can be written as
$$
 \delta \phi^{-1}_0 = -\sum_{\ell=-1}^\infty k^\ell
y^{\ell+1}.\eqno(47)
$$
Equation (47) is precisely the $w_{1+\infty}$ transformation rule of Eq.\
(3), after identifying the scalar field $\varphi$ of Eq.\ (3) with the
field $(-\phi^{-1}_0)$ here.

Substituting the relations (44) back into Eq.\ (34),  we find that the
$\delta x^+$ terms as well
as the terms proportional to $\delta \phi^{\{-1\}}$ on the left-hand side
and to $\gamma^{\{-1\}}$ on the right-hand side all cancel. We are left
with the result
$$
  \eqalign{
          &\cdots e^{\phi^{\{4\}}}\cdots e^{\phi^{\{2\}}}
            \bigg({1-e^{\phi^{\{1\}}}\over \phi^{\{1\}}}\bigg)\wedge
\delta\phi^{\{1\}}
       +\cdots e^{\phi^{\{4\}}} e^{\phi^{\{3\}}} \bigg({1-e^{\phi^{\{2\}}}
\over \phi^{\{2\}}}\bigg)\wedge
        \delta\phi^{\{2\}}\cr
        &\phantom{\cdots e^{\phi^{\{4\}}}\cdots e}
+\cdots e^{\phi^{\{4\}}}\bigg({1-e^{\phi^{\{3\}}}\over \phi^{\{3\}}}\bigg)
\wedge \delta\phi^{\{3\}}
         +\cdots\cr
&\phantom{\cdots e^{\phi^{\{4\}}}\cdots e^{\phi^{\{2\}}}
            \bigg(1-}= \cdots e^{\phi^{\{4\}}}\cdots
e^{\phi^{\{2\}}}e^{\phi^{\{1\}}}\wedge
        (\gamma^{\{0\}}+\gamma^{\{1\}}+\gamma^{\{2\}}\cdots)
-\gamma^{\{0\}}.\cr} \eqno(48)
$$
 From this formula, we can read off the transformation rules for
$\phi^{\{2N+1\}}$ and $\phi^{\{2N\}}$ as follows:
$$
\eqalign{
   \delta\phi^{\{2N+1\}}&=
-\gamma^{\{2N+1\}}-e^{\phi^{\{1\}}}\wedge\gamma^{\{2N\}}-
e^{\phi^{\{2\}}}e^{\phi^{\{1\}}} \wedge\gamma^{\{2N-1\}}\cr
&-\cdots-e^{\phi^{\{2N+1\}}}\cdots e^{\phi^{\{1\}}} \wedge
\gamma^{\{0\}}+e^{\phi^{\{N+1\}}}\bigg({1-e^{\phi^{\{N\}}}\over
\phi^{\{N\}}}\bigg)\wedge \delta \phi^{\{N\}}\cr
&\phantom{-\cdots-}+e^{\phi^{\{N+2\}}}e^{\phi^{\{N+1\}}}
e^{\phi^{\{N\}}}
\bigg({1-e^{\phi^{\{N-1\}}}\over
\phi^{\{N-1\}}}\bigg)\wedge \delta \phi^{\{N-1\}}\cr
&\phantom{-\cdots--\cdots-}+\cdots +e^{\phi^{\{2N\}}}\cdots
e^{\phi^{\{2\}}}\bigg({1-e^{\phi^{\{1\}}} \over \phi^{\{1\}}}\bigg)\wedge
\delta \phi^{\{1\}},\cr}\eqno(49)
$$
where $N=0,1,2,...$ and only terms with upper indices summing to $2N+1$ and
parameters $\gamma^{\{0\}}$, $\gamma^{\{1\}}$,$\ldots$, $\gamma^{\{2N+1\}}$
are to be kept. Similarly, for the transformation rule for $\phi^{\{2N\}}$ we
find the result
$$
\eqalign{
    \delta \phi^{\{2N\}} &=
-\gamma^{\{2N\}}-e^{\phi^{\{1\}}}\wedge\gamma^{\{2N-1\}}-
e^{\phi^{\{2\}}}e^{\phi^{\{1\}}} \wedge\gamma^{\{2N-2\}}\cr
&-\cdots-e^{\phi^{\{2N\}}}\cdots e^{\phi^{\{1\}}} \wedge \gamma^{\{0\}}
+\bigg({1-e^{\phi^{\{N\}}}\over \phi^{\{N\}}}\bigg)\wedge \delta
\phi^{\{N\}}\cr
&\phantom{-\cdots-}
+e^{\phi^{\{N+1\}}}e^{\phi^{\{N\}}}\bigg({1-e^{\phi^{\{N-1\}}}\over
\phi^{\{N-1\}}} \bigg)\wedge \delta \phi^{\{N-1\}}\cr
&\phantom{-\gamma^{\{2N\}}-e^{\phi^{\{1\}}}}
+\cdots+e^{\phi^{\{2N-1\}}}\cdots
e^{\phi^{\{2\}}}\bigg({1-e^{\phi^{\{1\}}}\over \phi^{\{1\}}}\bigg)\wedge\delta
\phi^{\{1\}},\cr}\eqno(50)
$$
where $N=1,2,3,...$ and only terms with upper indices summing to
$2N$ and parameters $\gamma^{\{0\}}, \gamma^{\{1\}},..., \gamma^{\{2N\}}$
are to be kept.

We next consider a number of examples that illustrate the use of these
formulas and give us the results for the low-lying levels.  At level
$\ell=1$, (49) yields
$$
\delta \phi^{\{1\}} =-\gamma^{\{1\}}-\phi^{\{1\}}\wedge \gamma^{\{0\}}.
\eqno(51
)
$$
Restricting attention to the independent field $\phi^1_{-2}$ and using
the commutation rules (2), we find from (51)
$$
    \delta \phi^1_{-2}=-\gamma^1_{-2}-2\phi^1_{-2}\partial_+\gamma^0_{-1}+
                    \partial_+\phi^1_{-2}\gamma^0_{-1}. \eqno(52)
$$
At level $\ell=2$, from (50) we find
$$
\delta \phi^{\{2\}} =-\gamma^{\{2\}}-{1\over
2}\phi^{\{1\}}\wedge\gamma^{\{1\}}-\phi^{\{2\}}\wedge\gamma^{\{0\}},
\eqno(53
)
$$
where we have used $\delta\phi^{\{1\}}$ as found in (51). Again, restricting
this to the left-edge field $\phi^2_{-3}$ and using the commutation
relations (2), we find
$$
 \delta\phi^2_{-3}=-\gamma^2_{-3}+\gamma^0_{-1}\partial_+\phi^2_{-3}
-3\partial_+
  \gamma^0_{-1}\phi^2_{-3} +\gamma^1_{-2}\partial_+\phi^1_{-2}
-\partial_+\gamma^1_{-2}\phi^1_{-2}.       \eqno(54)
$$
Note that only fields and parameters corresponding to the left edge of the
Fig.\ 1 diagram occur in these results. For the next two levels the results
are
$$
\eqalign{
  \delta\phi^{\{3\}} &=
-\gamma^{\{3\}}-\phi^{\{1\}}\wedge\gamma^{\{2\}}-\ft16(\phi^{\{1\}})^2
 \wedge \gamma^{\{1\}}-\phi^{\{3\}}\wedge \gamma^{\{0\}}, \cr
   \delta\phi^{\{4\}} &= -\gamma^{\{4\}}-\phi^{\{1\}}\wedge\gamma^{\{3\}}
-\ft12
   \big(\phi^{\{2\}}+(\phi^{\{1\}})^2\big)\wedge\gamma^{\{2\}}\cr
&\phantom{\delta\phi^{\{4\}}}-\ft14\big(\phi^{\{2\}}+\ft12(\phi^{\{1\}})^2\big
)
\phi^{\{1\}}\wedge
\gamma^{\{1\}}-\phi^{\{4\}}\wedge\gamma^{\{0\}}.\cr} \eqno(55)
$$
\np
\noindent{\bf 5. Other coset realisations}
\bigskip
     The possibility of eliminating most of the original $\phi^\ell_m$
Goldstone fields by the covariant constraints (24), thus reducing the
essential set of Goldstone fields to just the $\phi^\ell_{-\ell-1}$,
suggests that there should be an alternative coset space construction
giving these fields directly as the only coset representatives.  We shall
restrict ourselves here to cosets formed from the generators of
$w^\uparrow_{1+\infty}$ as shown in Fig.\ 1. In order to obtain only
coset parameters corresponding to the left edge of Fig.\ 1, one needs to
divide out by a group corresponding to the complement of the left-edge
generators in $w^\uparrow_{1+\infty}$.  This can be done because these
generators, $\{v^\ell_{-\ell}; \ell\ge-1\}$, close amongst themselves to
form a subalgebra which we shall denote $w^+_{1+\infty}$, in analogy to the
Virasoro subalgebra ${\rm Vir}^+$.  The coset
$w^\uparrow_{1+\infty}/w^+_{1+\infty}$ so obtained provides an alternative
non-reductive construction of the Goldstone-field transformation rules of
section 4.  The structure of this coset construction is summarised in Fig.\
2, using the same notation for the coset ($\circ$), divisor subgroup ($\ds$)
and spatial coordinate $x^+$ ($\bullet$) generators as in Fig.\ 1: \bigskip
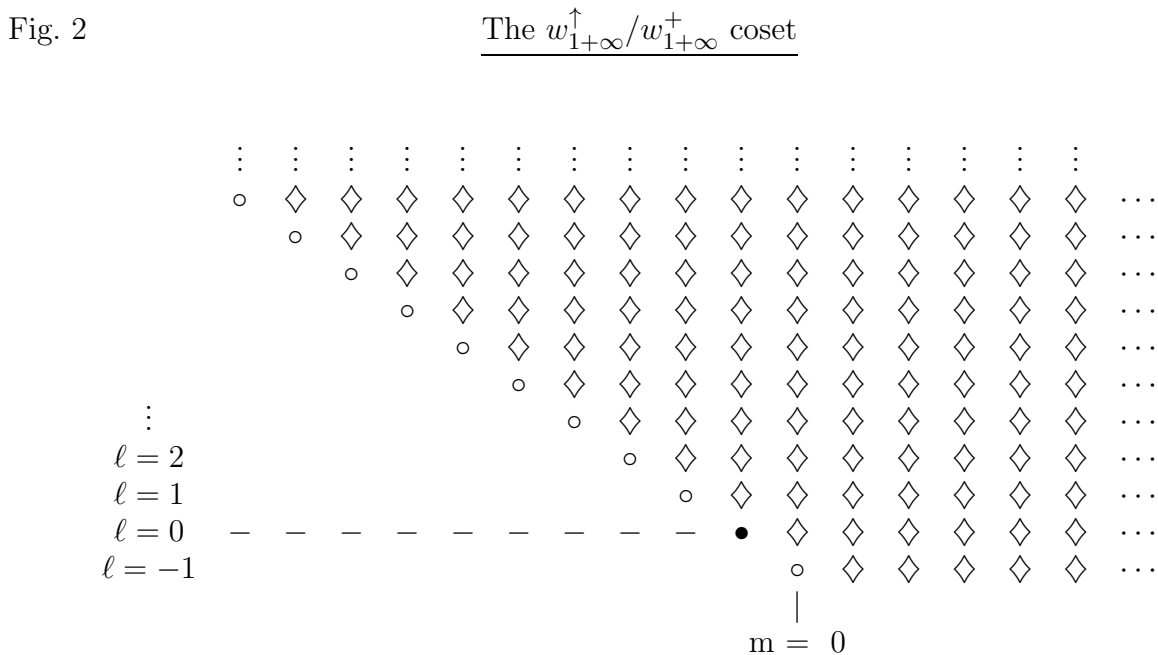
\leftline{Fig.\ 2\hskip 5.25cm\underscore{The
$w^\uparrow_{1+\infty}/w^+_{1+\infty}$ coset}} \nobreak
$$
\matrix{\px&\vdots&\vdots&\vdots&\vdots&\vdots&\vdots&\vdots&
\vdots&\vdots&\vdots&\vdots&\vdots&\vdots&\vdots&\vdots&\vdots&\cr
\px&\circ&\ds&\ds&\ds&\ds&\ds&\ds&\ds&\ds&\ds&\ds&\ds&\ds&\ds&\ds&\ds&\cdots
\cr
\px&&\circ&\ds&\ds&\ds&\ds&\ds&\ds&\ds&\ds&\ds&\ds&\ds&\ds&\ds&\ds&\cdots\cr
\px&&&\circ&\ds&\ds&\ds&\ds&\ds&\ds&\ds&\ds&\ds&\ds&\ds&\ds&\ds&\cdots\cr
\px&&&&\circ&\ds&\ds&\ds&\ds&\ds&\ds&\ds&\ds&\ds&\ds&\ds&\ds&\cdots\cr
\px&&&&&\circ&\ds&\ds&\ds&\ds&\ds&\ds&\ds&\ds&\ds&\ds&\ds&\cdots\cr
\px&&&&&&\circ&\ds&\ds&\ds&\ds&\ds&\ds&\ds&\ds&\ds&\ds&\cdots\cr
\smash{\vdots}&&&&&&&\circ&\ds&\ds&\ds&\ds&\ds&\ds&\ds&\ds&\ds&\cdots\cr
\ell=2&&&&&&&&\circ&\ds&\ds&\ds&\ds&\ds&\ds&\ds&\ds&\cdots\cr
\ell=1&&&&&&&&&\circ&\ds&\ds&\ds&\ds&\ds&\ds&\ds&\cdots\cr
\ell=0&-&-&-&-&-&-&-&-&-&\bullet&\ds&\ds&\ds&\ds&\ds&\ds&\cdots\cr
\ell=-1&&&&&&&&&&&\circ&\ds&\ds&\ds&\ds&\ds&\cdots\cr
&&&&&&&&&&&\vert&&&&&&\cr
&&&&&&&&&&\px\rlap{m}&=&\llap{0}\px&&&&&\cr}
$$

     Yet other coset realisations may be constructed by observing, from Eq.\
(2), that one may also form closed subalgebras of $w^\uparrow_{1+\infty}$
by transferring all but a finite number of left-edge generators into the
divisor, which then is generated
by $\big\{w^+_{1+\infty}\oplus\{v^\ell_{-\ell-1}; \ell> N\}\big\}$.  The
resulting diagram of generators is shown in Fig.\ 3.  The finite-dimensional
Goldstone field realisation corresponding to Fig.\  3 can also be extracted
from the explicit transformation rules for the Goldstone fields given in
section 4.  The transformation rules (49, 50) have a ``triangular''
structure: each left-edge Goldstone field of our surviving set transforms
into itself and into fields lower down on the left edge of the  Fig.\ 1
diagram, but not into fields higher up on the left edge.  Consequently, it
is possible to consistently truncate our infinite set of left-edge
Goldstone  fields down to the finite set $(\phi^\ell_{-\ell-1};
\ell=-1,1,2,\ldots,N)$.  This finite set then corresponds precisely to the
Goldstone fields of the non-reductive coset of Fig.\ 3.
\bigskip
\leftline{Fig.\ 3\hskip 5cm\underscore{A finite-dimensional coset}}
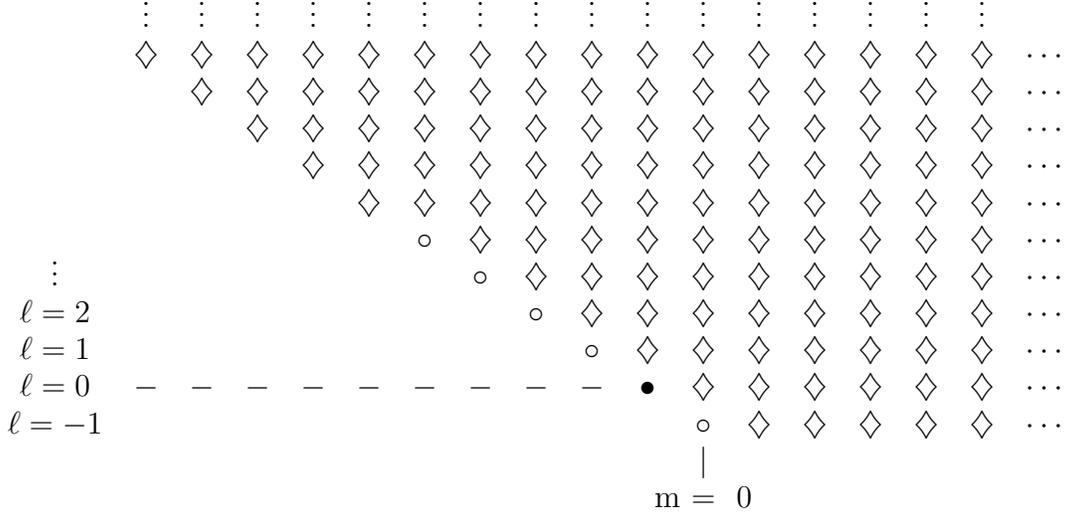
\nobreak
$$
\matrix{\px&\vdots&\vdots&\vdots&\vdots&\vdots&\vdots&\vdots&
\vdots&\vdots&\vdots&\vdots&\vdots&\vdots&\vdots&\vdots&\vdots&\cr
\px&\ds&\ds&\ds&\ds&\ds&\ds&\ds&\ds&\ds&\ds&\ds&\ds&\ds&\ds&\ds&\ds&\cdots\cr
    \px&&\ds&\ds&\ds&\ds&\ds&\ds&\ds&\ds&\ds&\ds&\ds&\ds&\ds&\ds&\ds&\cdots\cr
      \px&&&\ds&\ds&\ds&\ds&\ds&\ds&\ds&\ds&\ds&\ds&\ds&\ds&\ds&\ds&\cdots\cr
        \px&&&&\ds&\ds&\ds&\ds&\ds&\ds&\ds&\ds&\ds&\ds&\ds&\ds&\ds&\cdots\cr
          \px&&&&&\ds&\ds&\ds&\ds&\ds&\ds&\ds&\ds&\ds&\ds&\ds&\ds&\cdots\cr
            \px&&&&&&\circ&\ds&\ds&\ds&\ds&\ds&\ds&\ds&\ds&\ds&\ds&\cdots\cr
   \smash{\vdots}&&&&&&&\circ&\ds&\ds&\ds&\ds&\ds&\ds&\ds&\ds&\ds&\cdots\cr
             \ell=2&&&&&&&&\circ&\ds&\ds&\ds&\ds&\ds&\ds&\ds&\ds&\cdots\cr
               \ell=1&&&&&&&&&\circ&\ds&\ds&\ds&\ds&\ds&\ds&\ds&\cdots\cr
\ell=0&-&-&-&-&-&-&-&-&-&\bullet&\ds&\ds&\ds&\ds&\ds&\ds&\cdots\cr
                  \ell=-1&&&&&&&&&&&\circ&\ds&\ds&\ds&\ds&\ds&\cdots\cr
                                         &&&&&&&&&&&\vert&&&&&&\cr
                       &&&&&&&&&&\px\rlap{m}&=&\llap{0}\px&&&&&\cr}
$$
\bigskip
\noindent{\bf 6. Invariant actions}
\bigskip
     The canonical way to construct an invariant action with a non-linearly
realised symmetry from a reductive coset-space construction is to use the
``vierbein'' and ``connection'' components of the Maurer-Cartan form to build
a
Goldstone-field-dependent Lagrangian that is locally invariant under the
divisor
group of the coset.  In the present case, however, the covariant
inverse-Higgs constraints (24) reduce the $(+)$ component of the
Maurer-Cartan form (16) completely down to the coordinate differential
$(-dx^+)$ --- with even the connection terms $\omega_{(+)\,m}^0$
eliminated.  Any Lagrangian must have an overall $d=2$ Lorentz weight equal
to zero and therefore must involve both $\partial_+$ and $\partial_-$
derivative terms.  Thus, the absence of any non-trivial $(+)$ components of
the Maurer-Cartan form means that one cannot use the canonical procedure to
construct actions.

     Nonetheless, if we relax the condition of strict invariance of the
Lagrangian and seek only an invariant action, {\it i.e.}\ if we allow the
Lagrangian to transform by a total derivative, then a simple action is ready
to hand --- constructed from the free scalar Lagrangian for the field
$\phi^{-1}_0$
alone:
$$
{\cal L}_0 = \ft12\partial_+\phi^{-1}_0\partial_-\phi^{-1}_0.\eqno(56)
$$
It may be verified that this Lagrangian transforms by a total derivative
under the full set of $w^\uparrow_{1+\infty}$ transformations (47):
$$
 \delta \phi^{-1}_0 = -\sum_{\ell=-1}^\infty k^\ell
(\partial_+\phi^{-1}_0)^{\ell+1}.\eqno(57)
$$

The field equation $\partial_-\partial_+\phi=0$ which follows from
(56) is covariant under (57) since it is equivalent to the manifestly
covariant set of equations $E^{-1}_{(-)\,m\ge1}=0$, where
$E^{-1}_{(-)\,m}$ is given in (26), upon use of the set of manifestly
covariant inverse Higgs constraints (21).\footnote{$^\ast$}{We recall that
the set of the $E^{-1}_{(-)\,m}$ is manifestly covariant since it is
manifestly covariant under the ${\rm Vir}^+$ divisor subgroup, and hence
transforms homogeneously under the full $w^\uparrow_{1+\infty}$.  We note
also from (27, 28) that the dynamical equations $E^{-1}_{(-)\,m\ge1}=0$ have
the property of setting to zero both $E^0_{(-)\,-1}$ and
$\omega^0_{(-)\,m}$.} By contrast, the existence of the invariant action
$\int d^2x{\cal L}_0$ and its relation to the general scheme of
$w^\uparrow_{1+\infty}$ nonlinear realisations that we have been presenting
remains something of a mystery. The existence of the action requires firstly
a single-Goldstone-scalar realisation of $w^\uparrow_{1+\infty}$ symmetry
that we can obtain as the $N=0$ finite-component realisation of section 5.
Even given this, however, we do not have a canonical method of constructing
an action,  so the existence of (56) is something of an accident.  It is
worth recalling again the analogy of $d=4$ conformal symmetry that we
developed in section 2.  The existence of classical conformally-invariant
theories such as Yang-Mills theory is also in a sense accidental.  We saw
that nonlinear realisations of $SO(4,2)$ would generically require coupling
to the dilaton Goldstone field $\varphi(x)$.  Only in special cases,
such as that of Yang-Mills theory, is there a gauge symmetry that causes this
field to decouple, leaving a realisation on the minimal coset space for
conformal symmetry, {\it i.e.}\ on compactified Minkowski space ${\cal
M}^\sharp=SO(4,2)/(P'\times\{D\})$.

     In the present case, one may view the existence of the single-scalar
Lagrangian (56) as a consequence of similar gauge symmetries, in this case
for all the Goldstone fields $(\phi^\ell_{-\ell-1}; \ell\ge1)$ that do not
appear in (56).  As in the Yang-Mills case, such gauge
symmetries at the classical level might be suspected to be vulnerable to
anomalies in which the classically-decoupled Goldstone fields could
re-couple to the theory at the quantum level.  Indeed, in $w_\infty$-gravity
theory, which has analogous gauge symmetries, it appears that there
are in fact such anomalies.  Owing to the factorisation of the classical
currents $(\partial_+\phi)^{\ell+2}$, the $w_\infty$-gravity Lagrangian has
an infinite set of ``Stueckelberg'' symmetries that allow one to gauge away
all but the $\ell=0$ gauge field (when coupling to a single scalar as we
are considering here) [1].  At the quantum level, one may arrange for finite
renormalisations of the $w_\infty$ currents so as to preserve their
symmetries at the quantum level [10] (albeit at the price of having the
algebra be renormalised from $w_\infty$ to $W_\infty$), but the Stueckelberg
symmetries for the $\ell\ge1$ gauge fields appear to become anomalous in the
process.

     Free second-order scalar actions involving the higher left-edge Goldstone
fields $\phi^\ell_{-\ell-1}$ cannot be made because they do not have Lorentz
weight zero, since the Lorentz weight of $\phi^\ell_{-\ell-1}$ is $-(\ell+1)$.
One may, however, restore a set of local gauge symmetries and at the
same time couple the higher Goldstone fields to currents built from
$\phi^{-1}_0$.  In order to couple the higher Goldstone fields in this way,
one may take the $w_{\infty}$-gravity action of Ref.\ [1] but with its
fundamental gauge fields replaced by composite fields constructed so as to
transform correctly according to standard gauge field transformation rules
under the $w^\uparrow_\infty$ transformations.

     Since we have not introduced Goldstone fields for
any of the Virasoro generators (not counting the coordinate $x^+$, which is
the coset parameter for $v^0_{-1}$, but is not a field), there is no way for
us to introduce a composite gauge field for the Virasoro generators
$v^0_m$.\footnote{$^\ast$}{In this paper, we have chosen to work with a
linear realisation of the ${\rm Vir}^+$ algebra on Goldstone fields,
obtaining an essentially reductive sonstruction in the sense explained in
the introduction.  An alternative procedure would be to take the
$v^0_{m\ge1}$ generators into the coset instead of leaving them in the
divisor subalgebra.  In that case, the divisor would be generated by
$L_0=v^0_0$ alone and the construction would be strictly reductive, even
for the coodinate $x^+\leftrightarrow v^0_{-1}$.  Such a formalism could be
obtained from that of the present paper by ``unfixing'' a local $v^0_1$
symmetry in the action and transformation rules.  This would include an
$A^0$ composite gauge field among those appearing in (59).  In the
coset constructions, the equivalent modifications would be obtained by
including a factor $e^{-\phi^{\{0\}}}$ to the extreme {\it right} of all
the other coset elements in (14); by placing this factor in this position,
one avoids upsetting the covariance of the inverse Higgs constraints (25)
for the $\ell\ne0$ levels.  Such a construction may be found in Ref.\ [7].}
Moreover, the $w_\infty$ gravity action of Ref.\ [1] has a local $w_\infty$
symmetry but not a full local $w_{1+\infty}$ symmetry, and consequently does
not contain a gauge field for the $v^{-1}_m$ generators (in fact, no
gauge-invariant construction of this type with full local $w_{1+\infty}$
symmetry exists).  Accordingly, we shall construct here an action with
composite gauge fields for the $v^{\ell\ge1}_m$ generators only.  Truncation
of the full set of gauge fields $\{A^\ell(x^+,x^-)\}$ for
$w^\uparrow_{1+\infty}$ to the set $\{A^{\ell\ge1}\}$ is consistent for
a concurrent restriction of the symmetry group to $w_\infty$, as one may
verify directly from the transformation rules for gauge fields [1], under
local transformations with parameters $k^\ell(x^+,x^-)$ as given in (45),
but now allowed to be $x^-$-dependent also:
$$
\delta
A^\ell=\partial_-k^\ell-\sum_{j=0}^{\ell+1}[(j+1)A^j\partial_+k^{\ell-j} -
(\ell-j+1)k^{\ell-j}\partial_+A^j].\eqno(58)
$$

Given a set of composite gauge fields $A^\ell(\phi(x^+,x^-))$ that
transform according to (58), one can then take over the form of the
$w_\infty$-gravity Lagrangian from [1], but restricted here to gauge fields
with $\ell\ge1$:
$$
{\cal L}_{\infty}={\cal L}_0-\sum_{\ell=1}^\infty{1\over
\ell+2}A^\ell(\phi)(-\partial_+\phi^{-1}_0)^{\ell+2}.\eqno(59)
$$
Given (58), this action will be invariant under global
({\it i.e.}\ $x^-$-independent) $w^\uparrow_\infty$ transformations.  From
its origins in $w_\infty$-gravity, one knows that (59) is in fact also
invariant under {\it local} $k^\ell(x^+,x^-)$ transformations for
$\ell\ge1$.  Using these, one may of course gauge away to zero all of the
higher left-edge Goldstone fields $\phi^\ell_{-\ell-1}$, and hence eliminate
all of the ``pure-gauge'' composite gauge fields
$A^{\ell\ge1}(\phi(x^+,x^-))$.  In this special gauge, the Lagrangian (58)
must thus reduce back to the free-field Lagrangian (56) for $\phi^{-1}_0$
alone.  This is the sense in which local gauge symmetries cause the
Goldstone fields $\phi^\ell_{-\ell-1}$ to decouple classically, just as
local scale invariance causes the dilaton to decouple from classical
Yang-Mills theory.  The existence of these local gauge symmetries also gives
us an alternative way to obtain the form of the Lagrangian (59): one may
equivalently start from (56) and perform local $k^{\ell\ge1}(x^+,x^-)$
transformations, under which the free-field Lagrangian (56) is {\it not}
invariant, and then turn around and promote the parameters of these local
transformations to Goldstone fields $\phi^\ell_{-\ell-1}(x^+,x^-)$.

In order to obtain the explicit forms of the composite gauge fields appearing
in (59), we need to solve for the $A^\ell$ in terms of the $\phi^\ell_m$.
Following the work of Ref.\ [6], this may be done by generalising the
construction of the Maurer-Cartan form (16) through the
incorporation of a set of gauge fields $\{A^\ell_m(x^-)\}$.  Note that
by a ``dressing'' procedure analogous to (36, 37, 45), this
doubly-indexed set of functions of a single variable is equivalent to the
singly-indexed set of functions of two variables $A^\ell(x^+,x^-)$.  As
mentioned above, we introduce here gauge fields only for $\ell\ge1$ and
concurrently restrict the symmetry group to $w^\uparrow_\infty$.  With the
gauge fields included, the Maurer-Cartan form becomes
$$
\hat{\cal
P}=k^{-1}(d+A)k= \sum_{m\ge 0}E^{-1}_mv^{-1}_m +E^0_{-1}v^0_{-1}+
\sum_{\ell\ge 1 \atop m\ge -\ell-1}\hat E^\ell_mv^\ell_m + \sum_{m\ge
0}\hat\omega^0_m v^0_m,  \eqno(60)
$$
where $A=dx^-\sum_{\ell,m}A^\ell_m(x^-)v^\ell_m$ and the modified components
$\hat E^{\ell\ge1}_m$ and $\hat\omega^0_m$ now include
contributions from the gauge fields.  Since the $\hat E^{\ell\ge1}_m$
belong to linear representations of the coset divisor group ${\rm Vir}^+$,
they will transform homogeneously under global $w^\uparrow_\infty$
transformations (in fact, they also transform homogeneously under local
transformations $k^{\ell\ge1}_m$).  Thus, we may impose the covariant
conditions $$
\hat E^{\ell\ge1}_m\for_{x^+=0}=0\eqno(61)
$$
and these may then be solved to obtain the $A^{\ell\ge1}_m(x^-)$.  Note
that since ${\rm Vir}^+$ is the stability group of the point $x^+=0$, the
evaluation of (61) at this point is a covariant procedure.

The solution to the set of covariant constraints (61) is obtained,
following Ref.\ [6], by first writing the transformation rules (49, 50) for
the Goldstone fields $\phi^\ell_m(x^+=0,x^-)$ as
$$
\delta\phi^\ell_m\for_{x^+=0}=-\sum_{p,n}\left({\cal
F}^{\ell\ p}_{m\,n}(\phi)\alpha^p_n\right)\for_{x^+=0}, \eqno(62)
$$
where the nonsingular matrix ${\cal F}^{\ell\ p}_{m\,n}(\phi)$ is a nonlinear
functional of the $\phi^\ell_m$ (it may be seen to be nonsingular by noting
from (49, 50) that its leading term is the unit matrix $\delta^{\ell
p}\delta_{mn}$).  The solution to (61) is then given by
$$
A^\ell_m(\phi(x^-))=-\sum_{p,n}\left({\cal
F}^{(-1)}{}^{\ell\ p}_{m\,n}\partial_-\phi^p_n\right)\for_{x^+=0},
\qquad{\ell,\ p\ge1\atop m\ge-\ell-1;\ n\ge-p-1,}\eqno(63)
$$
where ${\cal F}^{(-1)}{}^{\ell\ p}_{m\,n}$ is the matrix inverse of
${\cal F}^{\ell\ p}_{m\,n}$, whose existence is guaranteed by the
nonsingularity of the latter.\footnote{$^\ast$}{The matrix inverse ${\cal
F}^{(-1)}{}^{\ell\ p}_{m\,n}$ in (63) may be taken simply with respect to the
$\ell,\ p\ge1$ submatrix of the full $w^\uparrow_{1+\infty}$ realisation
(the full realisation includes also $\ell,\ p=-1$) by virtue of the
``triangular'' nature of the realisation.  This has the consequence that the
$\ell,\ p\ge1$ submatrix of ${\cal F}^{(-1)}$ in the full realisation is
identical to the inverse of the $\ell,\ p\ge1$ submatrix of ${\cal
F}^{\ell\ p}_{m\,n}$.}  After obtaining the $A^\ell_m(\phi(x^-))$ from (63),
the $x^+$-dependent gauge fields $A^\ell(x^+,x^-)$ are then constructed,
analogously to (36, 37, 45), by dressing with $x^+$,
$$
A^\ell(x^+,x^-)=\sum_{m=0}^\infty(-x^+)^mA^\ell_{-\ell-1+m}(x^-).\eqno(64)
$$

One may verify directly that the composite $A^\ell(x^+,x^-)$ constructed in
this way do transform correctly according to (58) simply by varying, taking
into account the fact that (62) forms a realisation
of the $w^\uparrow_\infty$ algebra, as derived in section 4.  The
resulting Lagrangian (59) involves sums over infinite numbers of Goldstone
fields for each of the composite gauge fields, as a result of the matrix
inversion in (63), but owing to the nonsingularity of ${\cal
F}^{\ell\ p}_{m\,n}$ the result can be evaluated to arbitrary order in the
$\phi^\ell_m$.
\bigskip
\noindent{\bf 7. Conclusions}
\bigskip
     We have derived the general pattern of nonlinear realisations of
$w_{1+\infty}$ symmetry.  An infinite number of Goldstone fields arises in the
general nonlinear realisation, corresponding to the generators
$v^\ell_{-\ell-1}$ lying on the left edge of the diagram of Fig.\ 1.  In
general, these may all be expected to be present when the set of Goldstone
fields is used to promote a Virasoro-invariant theory to one with
$w_{1+\infty}$ invariance.  For
the pure Goldstone-field Lagrangian, however, there is a classical decoupling
of
all the higher fields $\phi^{\ell}_{-\ell-1}$ for $\ell\ge1$, leaving just a
free
scalar action for $\phi^{-1}_0$.

     The derivation of the Goldstone fields' transformation laws (49, 50)
follows straightforwardly from the theory of nonlinear realisations when one
starts from the essentially reductive coset space construction
$w_{1+\infty}^\uparrow/{\rm
Vir}^+$.  An infinite set of covariant constraints (25) can subsequently be
imposed to eliminate the inessential Goldstone fields.  These constraints,
however, also remove from the $(+)$ component of the Cartan-Maurer form (7)
all combinations of Goldstone fields with which one would normally construct
actions by making locally-invariant constructions with respect to the
linearly-realised divisor group ${\rm Vir^+}$.  Thus, the constuction of
$w_{1+\infty}$-invariant actions on a proper group-theoretical basis
still remains to be systematised.  One possible approach to a geometrical
interpretation of the action would be to view it as a kind of
``Chern-Simons'' action for $w_{1+\infty}$.

     We have dealt in this paper only with realisations of a single chiral
copy of $w_{1+\infty}$.  Of course, the free scalar action (56) is invariant
under both a left-handed as well as a right-handed copy of $w_{1+\infty}$.
Under the right-handed $w_{1+\infty}$, the r\^oles of $x^+$ and $x^-$ as
spectator coordinate and coset parameter are reversed as compared to the
left-handed copy.  The simultaneous nonlinear realisation of both copies on
a single scalar field $\varphi(x^+,x^-)$ is another subject on which the
analysis of the present paper can be extended. A related problem is the
geometrical origin of the multi-scalar $w_\infty$ realisations employed in
the $w_\infty$ gravity constructions of Ref.\ [1].  It may prove to be
fruitful to view the extra scalars as Goldstone fields for off-diagonal
combinations of higher left and right $v^\ell_0$ generators in the fashion
of the $W_3$ construction of Ref.\ [11].
 Since one may also view the study of nonlinearly-realised global symmetries
as the study of pure-gauge connections, there should also be interesting
relations between the present group-theoretic framework and the geometry of
$w_\infty$ gravity.  For example, topological $w_\infty$ gravity has been
interpreted as a theory of flat connections for the $SL(\infty,\R)$ ``wedge''
subalgebra of $w_{1+\infty}$ [12].  In that case, there are also covariant
constraints that allow the elimination of all but the set of connections
corresponding to the ``left-edge'' generators in our Fig.\ 1,
suggesting a relation to the present work. Finally, one would also like to
have a better understanding of how the realisations described in this paper
carry over to the quantum level, especially given the known anomaly
structure of $w_\infty$ gravity.
\bigskip
\noindent{\bf Acknowledgements}
\bigskip
     K.S.S. would like to thank Professor V.I. Ogievetsky, Dr.\ A.S.
Galperin, Dr.\ E.A. Ivanov and Dr.\  S. Krivonos for very stimulating
discussions.  We would also like to thank Professor Abdus Salam, the
International Atomic Energy Agency and UNESCO for hospitality at  the
International Center for Theoretical Physics in Trieste, where this work was
initiated. K.S.S. would also like to thank the Department of Physics at
Texas A\&M University and the Joint Institute for Nuclear Research, Dubna for
their hospitality during the course of the work.
\np
\centerline{\bf REFERENCES}
\bigskip
\frenchspacing
\item{[1]} E. Bergshoeff, C.N. Pope, L.J. Romans, E. Sezgin, X. Shen and K.S.
Stelle,\nl{\sl Phys. Lett.} {\bf 243B} (1990) 350.
\item{[2]} S. Coleman, J. Wess and B. Zumino, {\sl Phys. Rev.} {\bf 177}
(1969) 2239;
\item{} C.L. Callan, Jr., S. Coleman, J. Wess and B. Zumino, {\sl Phys. Rev.}
{\bf 177} (1969) 2247;
\item{} C.J. Isham, {\sl Nuovo Cimento} {\bf 59A} (1969) 356.
\item{[3]} D.V. Volkov, {\sl Fiz. Elem. Chastits At. Yadra} {\bf 4} (1973) 3;
\item{} V.I. Ogievetsky, in {\it Proc. 10$^{\nineit th}$ Winter School of
Theoretical Physics in Karpacz, Vol. 1}\nl (Wroc\l aw, 1974).
\item{[4]} G. Mack and Abdus Salam, {\sl Ann. Phys. (N.Y.)} {\bf 53} (1969)
174.
\item{[5]} A.B. Borisov and V.I. Ogievetsky, {\sl Theor. Mat. Fiz.} {\bf 21}
(1974) 329.
\item{[6]} E.A. Ivanov and V.I. Ogievetsky, {\sl Theor. Mat. Fiz.} {\bf 25}
(1975) 164.
\item{[7]} E.A. Ivanov and S.O. Krivonos, {\sl Lett. Math. Phys.} {\bf 7}
(1983) 523;
\item{[8]} K. Schoutens, {\sl Nucl. Phys.} {\bf B292} (1987) 150.
\item{[9]} B. Zumino, {\sl Nucl. Phys.} {\bf B127} (1977) 189.
\item{[10]} E. Bergshoeff, P.S. Howe, C.N. Pope, E. Sezgin, X. Shen and K.S.
Stelle,\nl{\sl Nucl. Phys.} {\bf B363} (1991) 163.
\item{[11]} E. Ivanov, S. Krivonos and A. Pichugin, ``Nonlinear
realisations of $W_3$ symmetry'', preprint JINR E2-91-328.
\item{[12]} H. Lu, C.N. Pope and X. Shen, {\sl Nucl. Phys.} {\bf B366}
(1991)  95.
\bye